\documentstyle[12pt,epsf]{article}

\catcode`\@=11
\long\def\@makefntext#1{
\protect\noindent \hbox to 3.2pt {\hskip-.9pt  
$^{{\ninerm\@thefnmark}}$\hfil}#1\hfill}		%CAN BE USED 

\def\@makefnmark{\hbox to 0pt{$^{\@thefnmark}$\hss}}  %ORIGINAL 
	
\def\ps@myheadings{\let\@mkboth\@gobbletwo
\def\@oddhead{\hbox{}
\rightmark\hfil\ninerm\thepage}   
\def\@oddfoot{}\def\@evenhead{\ninerm\thepage\hfil
\leftmark\hbox{}}\def\@evenfoot{}
\def\sectionmark##1{}\def\subsectionmark##1{}}

\renewcommand{\thefootnote}{\fnsymbol{footnote}}

\newcounter{sectionc}\newcounter{subsectionc}\newcounter{subsubsectionc}
\renewcommand{\section}[1] {\vspace*{0.6cm}\addtocounter{sectionc}{1} 
\setcounter{subsectionc}{0}\setcounter{subsubsectionc}{0}\noindent 
	{\normalsize\bf\thesectionc. #1}\par\vspace*{0.4cm}}
\renewcommand{\subsection}[1] {\vspace*{0.6cm}\addtocounter{subsectionc}{1} 
	\setcounter{subsubsectionc}{0}\noindent 
	{\normalsize\it\thesectionc.\thesubsectionc. #1}\par\vspace*{0.4cm}}
\renewcommand{\subsubsection}[1]
{\vspace*{0.6cm}\addtocounter{subsubsectionc}{1}
	\noindent {\normalsize\rm\thesectionc.\thesubsectionc.\thesubsubsectionc. 
	#1}\par\vspace*{0.4cm}}

\newcounter{appendixc}
\newcounter{subappendixc}[appendixc]
\newcounter{subsubappendixc}[subappendixc]

\renewcommand{\appendix}[1] {\vspace*{0.6cm}
        \refstepcounter{appendixc}
        \setcounter{figure}{0}
        \setcounter{table}{0}
        \setcounter{equation}{0}
        \renewcommand{\thefigure}{\Alph{appendixc}.\arabic{figure}}
        \renewcommand{\thetable}{\Alph{appendixc}.\arabic{table}}
        \renewcommand{\theappendixc}{\Alph{appendixc}}
        \renewcommand{\theequation}{\Alph{appendixc}.\arabic{equation}}
        \noindent{\bf Appendix \theappendixc #1}\par\vspace*{0.4cm}}

\def\abstracts#1{{
	\centering{\begin{minipage}{12.2truecm}\footnotesize\baselineskip=12pt\noindent
	\centerline{\footnotesize ABSTRACT}\vspace*{0.3cm}
	\parindent=0pt #1
	\end{minipage}}\par}} 

\renewenvironment{thebibliography}[1]
	{\begin{list}{\arabic{enumi}.}
	{\usecounter{enumi}\setlength{\parsep}{0pt}
\setlength{\leftmargin 1.25cm}{\rightmargin 0pt}
	 \setlength{\itemsep}{0pt} \settowidth
	{\labelwidth}{#1.}\sloppy}}{\end{list}}

\newcounter{itemlistc}
\newcounter{romanlistc}
\newcounter{alphlistc}
\newcounter{arabiclistc}

\newcommand{\fcaption}[1]{
        \refstepcounter{figure}
        \setbox\@tempboxa = \hbox{\footnotesize Fig.~\thefigure. #1}
        \ifdim \wd\@tempboxa > 6in
           {\begin{center}
        \parbox{6in}{\footnotesize\baselineskip=12pt Fig.~\thefigure. #1}
            \end{center}}
        \else
             {\begin{center}
             {\footnotesize Fig.~\thefigure. #1}
              \end{center}}
        \fi}

\newcommand{\tcaption}[1]{
        \refstepcounter{table}
        \setbox\@tempboxa = \hbox{\footnotesize Table~\thetable. #1}
        \ifdim \wd\@tempboxa > 6in
           {\begin{center}
        \parbox{6in}{\footnotesize\baselineskip=12pt Table~\thetable. #1}
            \end{center}}
        \else
             {\begin{center}
             {\footnotesize Table~\thetable. #1}
              \end{center}}
        \fi}

\def\@citex[#1]#2{\if@filesw\immediate\write\@auxout
	{\string\citation{#2}}\fi
\def\@citea{}\@cite{\@for\@citeb:=#2\do
	{\@citea\def\@citea{,}\@ifundefined
	{b@\@citeb}{{\bf ?}\@warning
	{Citation `\@citeb' on page \thepage \space undefined}}
	{\csname b@\@citeb\endcsname}}}{#1}}

\newif\if@cghi
\def\cite{\@cghitrue\@ifnextchar [{\@tempswatrue
	\@citex}{\@tempswafalse\@citex[]}}
\def\citelow{\@cghifalse\@ifnextchar [{\@tempswatrue
	\@citex}{\@tempswafalse\@citex[]}}
\def\@cite#1#2{{$\null^{#1}$\if@tempswa\typeout
	{IJCGA warning: optional citation argument 
	ignored: `#2'} \fi}}

 1
 1
 1

\font\ninerm=cmr9

\begin{document}
\rightline{\vbox{\halign{&#\hfil\cr
&ANL-HEP-CP-96-51\cr
&June 21, 1996\cr}}}
\vspace{0.6in}
\centerline{\normalsize\bf CALCULATION OF THE CROSS SECTION}
\centerline{\normalsize\bf FOR TOP QUARK PRODUCTION
\footnote{Invited paper presented by E. L. Berger at the XIth Topical Workshop 
on Hadron Collider Physics, Abano Terme, Padova, Italy, May, 1996}}
\baselineskip=22pt

%\vfill
%\vspace*{0.6cm}
\centerline{\footnotesize EDMOND L. BERGER}
\baselineskip=13pt
\centerline{\footnotesize\it High Energy Physics Division, Argonne National
Laboratory, Argonne, IL 60439-4815, USA}
\centerline{\footnotesize E-mail: ELB@hep.anl.gov}
\vspace*{0.3cm}
\centerline{\footnotesize and}
\vspace*{0.3cm}
\centerline{\footnotesize HARRY CONTOPANAGOS}
\baselineskip=13pt
\centerline{\footnotesize\it High Energy Physics Division, Argonne National
Laboratory, Argonne, IL 60439-4815, USA}
\centerline{\footnotesize E-mail: CONTOPAN@hep.anl.gov}
%\vfill
\vspace*{0.9cm}
\abstracts{We summarize calculations of the cross section for top quark
production at hadron colliders within the context of perturbative quantum
chromodynamics, including resummation of the effects of initial-state
soft gluon radiation to all orders in the strong coupling strength.
In our approach we resum the universal leading-logarithm contributions, and
we restrict the calculation to the region of phase space that is
demonstrably perturbative.  We compare our approach with other methods.  We
present predictions of the physical cross section as a function of the top
quark mass in proton-antiproton reactions at center-of-mass energies of 1.8 and
2.0 TeV, and we discuss estimated uncertainties.}

%\vspace*{0.6cm}
\normalsize\baselineskip=15pt
\setcounter{footnote}{0}
\renewcommand{\thefootnote}{\alph{footnote}}
\section{Introduction and Motivation}
\pagestyle{plain}
In this report we summarize calculations carried out in perturbative 
quantum chromodynamics (QCD) of the inclusive cross section for the production 
of top quark-antiquark ($t\bar{t}$) pairs in hadron reactions\cite
{ref:laeneno,ref:edpapero,ref:edpapert,ref:catani}.  We begin with a 
discussion of the motivation for the inclusion of the effects of intial 
state soft gluon radiation to all orders in the QCD coupling strength, and we 
review the general formalism of resummation.  We outline the method and 
domain of applicability of perturbative resummation that we developed in the 
past year\cite{ref:edpapero,ref:edpapert}, and we contrast this approach with 
other methods\cite{ref:laeneno,ref:catani}.   We present predictions of the 
physical cross section as a function of the top quark mass in proton-antiproton 
reactions at center-of-mass energies of 1.8 and 2.0 TeV.  

In hadron interactions at collider energies, $t\bar{t}$ pair production 
proceeds through partonic hard-scattering processes involving initial-state
light quarks $q$ and gluons $g$.  In lowest-order QCD, 
${\cal O}(\alpha_s^2)$,  the two partonic subprocesses 
are $q + \bar{q} \rightarrow t + \bar{t}$ and $g + g \rightarrow t + \bar{t}$.  
Calculations of the cross section through next-to-leading order, 
${\cal O}(\alpha_s^3)$, involve gluonic radiative corrections to these 
lowest-order subprocesses as well as contributions from the $q + g$ initial 
state\cite{ref:dawson}.  A complete fixed-order calculation at order 
${\cal O}(\alpha_s^n), n \ge 4$ does not exist.

The physical cross section for each production channel is obtained through 
the factorization theorem,
\begin{equation}
\sigma_{ij}(S,m^2)={4m^2\over S}\int_0^{{S\over 4m^2}-1}d\eta\Phi_{ij}\biggl[
{4m^2\over S}(1+\eta),\mu^2\biggr]\hat\sigma_{ij}(\eta,m^2,\mu^2) .
\label{feleven}
\end{equation}
The square of the total hadronic center-of-mass energy is $S$, the square of 
the partonic center-of-mass energy is $s$, and $m$ denotes the top mass.  
The variable $\eta={s \over 4m^2} - 1$ measures the distance from the 
partonic threshold.  The indices $ij\in\{q\bar{q},gg\}$ denote the initial 
parton channel.  The partonic cross section 
$\hat\sigma_{ij}(\eta,m^2,\mu^2)$ is obtained commonly from fixed-order QCD
calculations\cite{ref:dawson}, or, as described here, from calculations
that go beyond fixed-order perturbation theory through the inclusion of 
gluon resummation\cite{ref:laeneno,ref:edpapero,ref:edpapert,ref:catani}.  
The parton flux is 
$\Phi_{ij}(y,\mu^2)=\int_y^1{dx\over x}f_{i/h_1}(x,\mu^2)f_{j/h_2}(y/x,\mu^2)$,
where $f_{i/h_1}(x,\mu^2)$ is the density of partons of type $i$ in hadron 
$h_1$.
Henceforth, we use the notation $\alpha(\mu)\equiv \alpha_s(\mu)/\pi$, where 
$\mu$ is the common renormalization/factorization scale of the problem.  
Unless otherwise specified, $\alpha\equiv \alpha(\mu=m)$.  The total physical 
cross section is obtained after incoherent addition of the contributions from 
the the $q\bar{q}$ and $gg$ production channels.  We ignore the small 
contribution from the $qg$ channel.

A comparison of the partonic cross section at next-to-leading order with its 
lowest-order value reveals that the ratio becomes very large in the 
near-threshold region, i.e., the ``$K$-factor" at the partonic level 
$\hat K(\eta)$ becomes very large as $\eta \rightarrow 0$.  An illustration of
this behavior may be seen in Fig.~7 of Ref.~[3].  This large ratio casts doubt 
on the reliability of simple fixed-order perturbation theory for physical 
processes for which the near-threshold region in the subenergy variable 
contributes significantly to the physical cross section.  Top quark production 
at the Fermilab Tevatron is one such process, because the top mass is 
relatively large compared to the energy available.  Other examples include 
the production of hadronic jets that carry large values of transverse momentum 
and the production of pairs of supersymmetric particles with large mass.  To 
obtain more reliable theoretical estimates of the cross section in 
perturbative QCD, it is important first to identify and isolate the terms that 
provide the large next-to-leading order enhancement and then to resum these 
effects to all orders in the strong coupling strength.  

\section{Gluon Radiation and Resummation}  
The origin of the large threshold enhancement may be traced to initial-state
gluonic radiative corrections to the lowest-order channels.  Consider the 
subprocess $i(k_1)+j(k_2)\rightarrow t(p_1)+{\bar t}(p_2)+g(k)$.  We define the
variable $z$ through the partonic invariants\cite{ref:laeneno}
\begin{equation}
s=(k_1+k_2)^2,\ t_1=(k_2-p_2)^2-m^2,\ u_1=(k_1-p_2)^2-m^2,\ 
(1-z)m^2= s+t_1+u_1.
\label{invariants}
\end{equation}
Alternatively, $(1-z) = {2k \cdot p_1 \over m^2}$.  In the limit that 
$z \rightarrow 1$, the radiated gluon carries zero momentum.  After cancellation
of soft singularities and factorization of collinear singularities in 
${\cal O}(\alpha^3)$, there are left-over integrable logarithmic contributions
to the cross section associated with initial-state gluon radiation.  These 
contributions are proportional to $\ln(1-z)$.  

The partonic cross section may be expressed generally as
\begin{equation}
\hat{\sigma}_{ij}(\eta,m^2)=\int_{z_{min}}^1
dz\biggl[1+{\cal H}_{ij}(z,\alpha)\biggr]\hat{\sigma}_{ij}'(\eta,m^2,z).
\label{bone}
\end{equation}
We work in the $\overline{\mbox{MS}}$ factorization scheme in which
the $q$, $\bar{q}$ and $g$ densities and the next-to-leading order
partonic cross sections are defined unambiguously.  The lower limit of 
integration, $z_{min} = 1-4(1+\eta)+4\sqrt{1+\eta}$, is set by kinematics.  
The derivative 
$\hat{\sigma}_{ij}'(\eta,m^2,z)=d(\hat{\sigma}_{ij}^{(0)}(\eta,m^2,z))/dz$,
and $\hat{\sigma}_{ij}^{(0)}$ is the lowest-order ${\cal O}(\alpha^2)$ 
partonic cross section expressed in terms of inelastic kinematic variables to 
account for the emitted radiation.

Keeping only the leading logarithmic contributions through 
${\cal O}(\alpha^3)$, we may write the 
total partonic cross section as
\begin{equation}
\hat\sigma_{ij}^{(0+1)}(\eta,m^2)=\int_{z_{min}}^1dz
\left\{1+\alpha 2 C_{ij}\ln^2(1-z)\right\}\hat \sigma'_{ij}(\eta,z,m^2)\ ,
\label{twelvep}
\end{equation}
where $C_{q\bar{q}}=C_F=4/3$ and $C_{gg}=C_A=3$. 
As is illustrated in Fig.~1(a), the leading logarithmic contribution, integrated
over the near-threshold region $1 \ge z \ge 0$, provides an excellent 
approximation to the exact full next-to-leading order physical cross section 
as a function of the top quark mass.  
Although an exact fixed-order ${\cal O}(\alpha^4)$ calculation of 
$t\bar{t}$ pair production does not
exist, we may invoke universality with massive lepton-pair production 
($l\bar{l}$), the Drell-Yan process, to generalize Eq.~(\ref{twelvep}) to 
higher order.  In the near-threshold region, the kernel becomes
\begin{equation}
{\cal H}^{(0+1+2)}_{ij}(z,\alpha) \simeq 2\alpha C_{ij} \ell n^2 (1-z)
+ \alpha^2\biggl[2C^2_{ij} \ell n^4 (1-z) - {4\over 3} C_{ij} b_2
\ell n^3 (1-z)\biggr] .
\label{btwo}
\end{equation}
The coefficient $b_2=(11C_A-2n_f)/12$, and the number of flavors $n_f=5$. The 
further enhancement of the physical cross section produced by the 
${\cal O}(\alpha^4)$ leading logarithmic terms in the near-threshold region 
is shown in Fig.~1.  At $m =$ 175 GeV, we compute the following ratios of the
physical cross sections in the leading logarithmic approximation:   
$\sigma_{ij}^{(0+1)}/\sigma_{ij}^{(0)} = $1.22, and
$\sigma_{ij}^{(0+1+2)}/\sigma_{ij}^{(0+1)} =$ 1.14.  These ratios show
that the near-threshold logarithms build up cross section in a worrisome 
fashion.  The ratios suggest that perturbation
theory is not converging to a stable prediction of the cross section.   

The goal of gluon resummation is to sum the series in $\alpha^n \ln^{2n}(1-z)$ 
to all orders in $\alpha$ in order to obtain a more defensible prediction.
This procedure has been studied extensively for the Drell-Yan 
process\cite{ref:stermano}.  In essentially all resummation procedures, the 
large logarithmic contributions are exponentiated into a function of the QCD 
running coupling strength, itself evaluated at a variable momentum scale that 
is a measure of the radiated gluon momentum.  For example, in the approach of
LSvN\cite{ref:laeneno}, the resummed partonic cross section is written as
\begin{equation}
\hat{\sigma}_{ij}^{R;IRC}(\eta,\mu_o)=\int_{z_{min}}^{1-(\mu_o/m)^3}
dz{\rm e}^{E_{ij}(z,m^2)}\hat{\sigma}_{ij}'(\eta,m^2,z),
\label{bthree}
\end{equation}
where the exponent 
\begin{equation}
E_{ij}(z,m^2) \propto C_{ij}\alpha((1-z)^{2/3}m^2)\ell n^2 (1-z).
\label{bfour}
\end{equation}

\begin{figure}
\vspace*{13pt}
%\leftline{\hfill\vbox{\hrule width 5cm height0.001pt}\hfill}
{\hskip2.0cm}\hbox{\epsfxsize8.6cm\epsffile{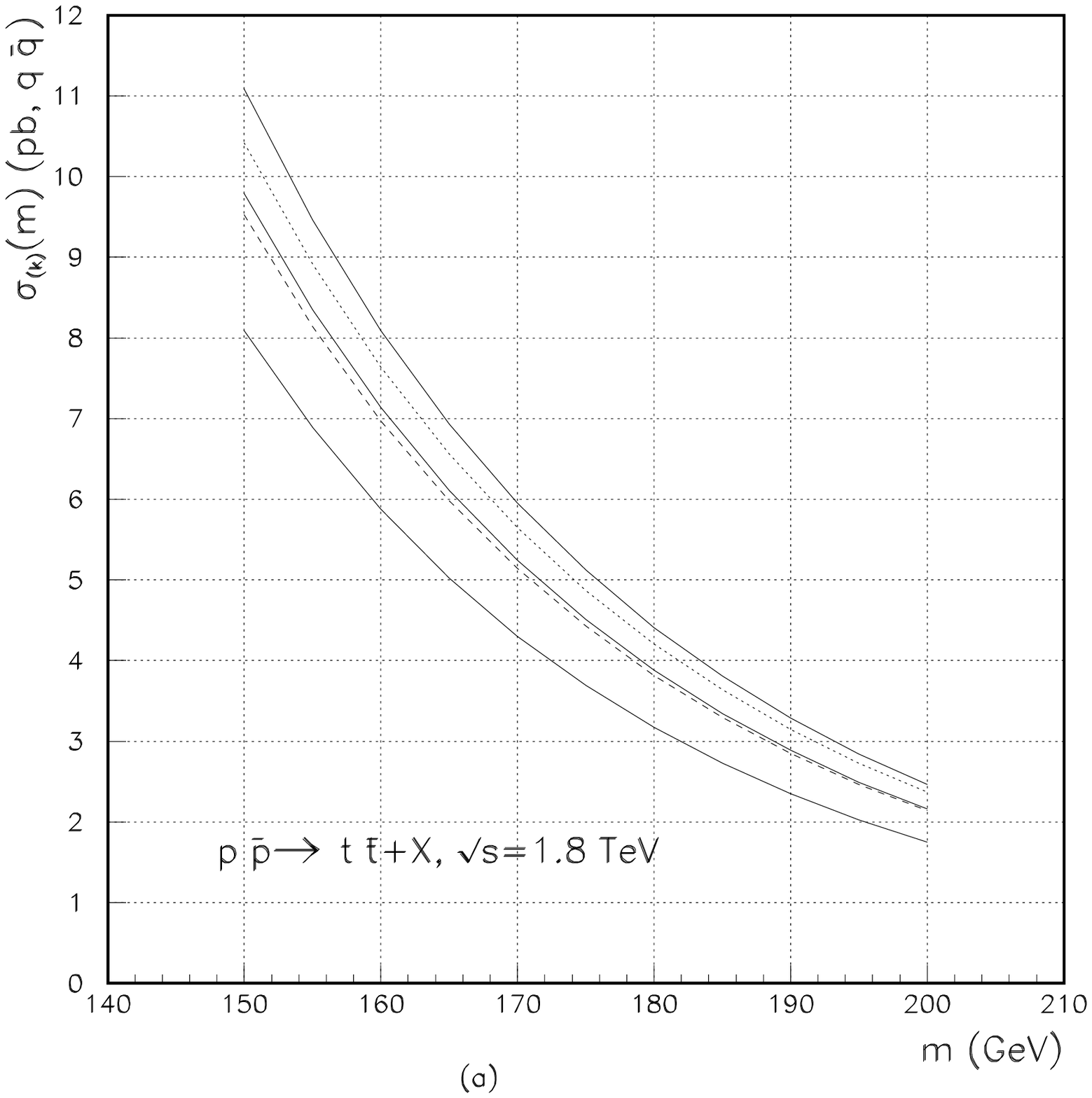}{\hskip -1.0cm}
\epsfxsize8.6cm\epsffile{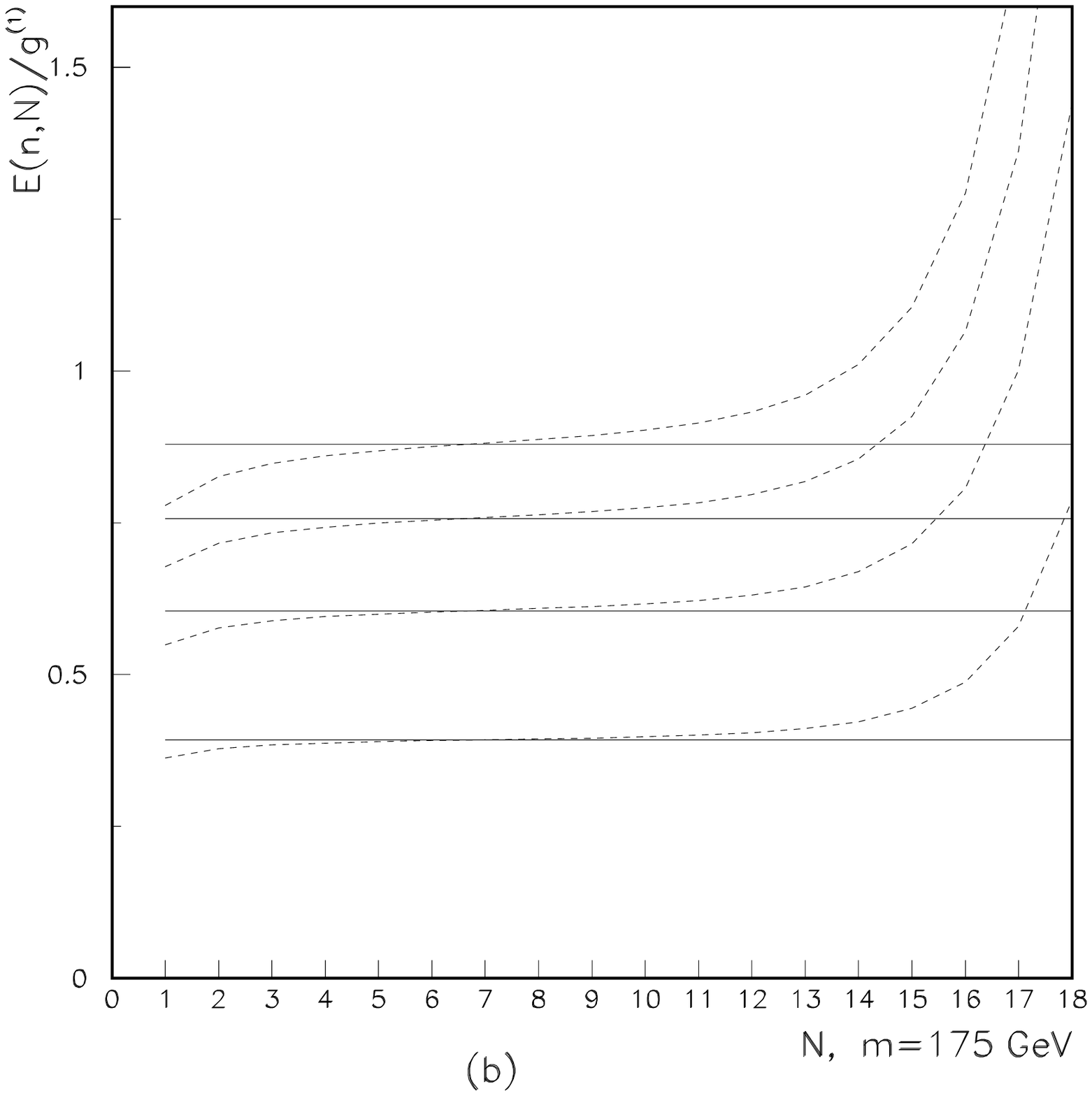}}
%\vspace*{1.4truein}             %ORIGINAL SIZE=1.6TRUEIN x 100% - 0.2TRUEIN
%\leftline{\hfill\vbox{\hrule width 5cm height0.001pt}\hfill}
\fcaption{(a) Physical cross sections in the $q\bar{q}$ channel in
the $\overline{{\rm MS}}$ scheme.
The solid lines denote the finite-order partial sums of the
universal leading-logarithmic contributions from the explicit
${\cal O}(\alpha^3)$ and ${\cal O}(\alpha^4)$ calculations for the
$t\bar{t}$ and Drell-Yan cross sections, respectively.
Lower solid: $\sigma^{(0)}$;
middle solid: $\sigma^{(0+1)}$;
upper solid: $\sigma^{(0+1+2)}$. The dashed curve represents the exact
next-to-leading order calculation for $t\bar{t}$ production,
in excellent agreement with $\sigma^{(0+1)}$. The dotted curve is our
resummed prediction. (b) Optimum number of perturbative terms in the exponent
with PVR  (solid family is for PVR, dashed for perturbative
approximation, both families increasing, for parametric values
$n=10,20,30,40$).}
\end{figure}
Different methods of resummation differ in theoretically and phenomenologically
important respects.  Formally, if not explicitly in some approaches, an 
integral over the radiated gluon momentum $z$ must be done over regions in 
which $z \rightarrow 0$.  Therefore, one significant distinction among methods 
has to do with how the inevitable ``non-perturbative" region is handled in each 
case.  Examination of Eqs.~(\ref{bthree}) and (\ref{bfour}) shows
that an infrared singularity is encountered in the soft-gluon limit
$z \rightarrow 1$:  owing to the logarithmic behavior of $\alpha(q^2)$,
$\alpha(q^2) \propto \ell n^{-1} (q^2/\Lambda_{QCD}^2)$,
$\alpha((1-z)^{2/3}m^2) \rightarrow \infty$ as $z \rightarrow 1$.  The infrared 
singularity is a manifestation of non-perturbative physics.  In the
approach of LSvN, this divergence of the integrand at the upper limit of 
integration necessitates introduction of the undetermined infrared  
cutoff (IRC) $\mu_o$ in Eq.~(\ref{bthree}), 
with $\Lambda_{QCD} \leq \mu_o \leq m$.
The cutoff prevents the integration over $z$ from reaching the Landau pole of 
the QCD running coupling constant.  The presence of an extra scale spoils the renormalization 
group properties of the overall expression.  The unfortunate dependence of the 
resummed cross section on this undetermined cutoff is important numerically 
since it appears in an exponent\cite{ref:laeneno}.  It is difficult to evaluate
theoretical uncertainties in a method that depends on an undetermined
infrared cutoff.  

\section{Perturbative Resummation}
The method of resummation we employ\cite{ref:edpapero,ref:edpapert} is
based on a perturbative truncation of principal-value resummation.  
The principal-value method (PVR)\cite{ref:stermano}
has an important technical advantage in that it does not depend on arbitrary
infrared cutoffs, as all Landau-pole singularities are by-passed by a Cauchy
principal-value prescription.  Because extra undetermined scales are absent,
the method also permits an evaluation of the perturbative regime of
applicability of the method, i.e., the region of the gluon radiation phase
space where perturbation theory should be valid.  

To illustrate how infrared cutoffs are avoided in the PVR method, it is useful 
to begin with an expression in moment ($n$) space for the exponent that 
resums the $\ell n (1-z)$ terms\cite{ref:stermanold}.  Factorization and 
evolution lead directly to exponentiation in moment space:  
\begin{equation}
E(n, m^2)= -\int\limits^1_0 dx {{x^{n-1}-1}\over{1-x}}
\int\limits^1_{(1-x)^2} {{d\lambda}\over{\lambda}} g
\left[ \alpha\left( \lambda m^2\right) \right].
\label{bsix}
\end{equation}
The function $g(\alpha)$ is calculable perturbatively, but the
behavior of $\alpha(\lambda m^2)$ leads to a divergence of the integrand when
$\lambda m^2 \rightarrow \Lambda_{QCD}^2$.  To tame the divergence, a cutoff
can be introduced in the integral over $x$ or directly in momentum space, in
the fashion of LSvN.  In the principal-value redefinition of resummation, the 
singularity is avoided by replacement of the integral over the real axis $x$ in
Eq.~(\ref{bsix}) by an integral in the complex plane along a contour $P$ that 
is symmetric under reflections across the real axis:
\begin{equation}
E^{PV}(n, m^2)\equiv -\int\limits_P
d\zeta {{\zeta^{n-1}-1}\over{1-\zeta}}
\int\limits^1_{(1-\zeta)^2}
{{d\lambda}\over{\lambda}}g\left[\alpha\left(\lambda m^2\right)\right].
\label{bseven}
\end{equation}
The function $E^{PV}(n, m^2)$ is finite since the Landau pole singularity
is by-passed.  In Eq.~(\ref{bseven}), all large soft-gluon threshold
contributions are included through the two-loop running of $\alpha$.

Equations~(\ref{bsix}) and~(\ref{bseven}) have identical perturbative 
content, but they have different non-perturbative content since the infrared 
region is treated differently in the two cases.  The non-perturbative content 
is not a prediction of perturbative QCD.  In our study 
of top quark production, we choose to use the exponent only in the 
region of phase space in which the perturbative content dominates.

We use the attractive finiteness of Eq.~(\ref{bseven}) to derive a 
perturbative asymptotic representation of $E(x,\alpha(m))$ that is 
valid in the moment-space interval
\begin{equation}
1<x\equiv \ln n< t\equiv {1\over 2\alpha b_2}.
\label{tseven}
\end{equation}
This perturbative asymptotic representation is
\begin{equation}
E_{ij}(x,\alpha)\simeq E_{ij}(x,\alpha,N(t))=
2C_{ij}\sum_{\rho=1}^{N(t)+1}\alpha^\rho
\sum_{j=0}^{\rho+1}s_{j,\rho}x^j\ .
\label{teight}
\end{equation}
Here
\begin{equation}
s_{j,\rho}=-b_2^{\rho-1}(-1)^{\rho+j}2^\rho c_{\rho+1-j}(\rho-1)!/j!\ ,
\label{tnine}
\end{equation}
and $\Gamma(1+z)=\sum_{k=0}^\infty c_k z^k$, where $\Gamma$ is the Euler gamma 
function.  
The number of perturbative terms $N(t)$ in Eq.~(\ref{teight}) is
obtained\cite{ref:edpapert} by optimizing the asymptotic approximation
$\bigg|E(x,\alpha)-E(x,\alpha,N(t))\bigg|={\rm minimum}$. 
Because of the range of validity in Eq.~(\ref{tseven}), terms in the exponent 
of the form $\alpha^k\ln^kn$ are of order unity, and terms with fewer powers
of logarithms, $\alpha^k\ln^{k-m}n$, are negligible.
This explains why resummation is completed in a finite number of steps. 
In addition, we discard monomials
$\alpha^k\ln^kn$ in the exponent because of the restricted leading-logarithm
universality between the $t\bar{t}$ and $l\bar{l}$ processes.

The exponent we use in our calculations of top quark production is
the truncation
\begin{equation}
E_{ij}(x,\alpha,N)=2C_{ij}\sum_{\rho=1}^{N(t)+1}\alpha^\rho s_\rho x^{\rho+1} ,
\label{tseventeen}
\end{equation}
with the coefficients
$s_\rho\equiv s_{\rho+1,\rho}=b_2^{\rho-1}2^\rho/\rho(\rho+1)$.  The number of
perturbative terms $N(t)$ is a function of only the top quark mass $m$.  This
expression contains no factorially-growing (renormalon) terms.  The 
perturbative region of phase space is far removed from the part of phase space
in which renormalons could be influential\cite{ref:catani}.  

In Fig.~1(b) we illustrate the validity of the asymptotic approximation 
for a value of t corresponding to $m=175$ GeV.  Optimization
works perfectly, with $N(t)=6$, 
and the plot demonstrates the typical breakdown of the asymptotic
approximation if $N$ were allowed to increase beyond $N(t)$. This  
is the exponential rise of the infrared (IR) renormalons,
the $(\rho-1)!$ growth in the second term of Eq.~(\ref{tnine}). 
As long as $n$ is in the interval of Eq.~(\ref{tseven}),
all the members of the family in $n$ are optimized 
at the same $N(t)$, showing that the optimum number of 
perturbative terms is a function of $t$, i.e., of $m$ only.

It is valuable to stress that we can derive the perturbative expressions,
Eqs.~(\ref{tseven}), (\ref{teight}), and (\ref{tnine}), from the unregulated
exponent Eq.~(\ref{bsix})
without the PVR prescription, although with less certitude.   We discuss
this point in some detail in our long paper\cite{ref:edpapert}.

After inversion of the Mellin transform from moment space to the physically 
relevant momentum space, the resummed partonic cross sections, 
including all large threshold corrections, can be written
in the form of Eq.~(\ref{bone}), but with Eq.~(\ref{btwo}) replaced by 
\begin{equation} 
{\cal H}^{R}_{ij}(z,\alpha)=\int_0^{\ln({1\over 1-z})}
dx{\rm e}^{E_{ij}(x,\alpha)}
\sum_{j=0}^\infty Q_j(x,\alpha)
\ .\label{bafour}
\end{equation}
The leading large threshold corrections are contained in the exponent 
$E_{ij}(x,\alpha)$, a calculable polynomial in $x$.  The functions 
$\{Q_j(x,\alpha)\}$ arise from the analytical inversion of the Mellin 
transform from moment space to momentum space. 
These functions are produced by the resummation and are expressed in 
terms of successive derivatives of $E$:
$P_k(x,\alpha)\equiv\partial^k E(x,\alpha)/k! \partial^k x$.  Each 
$Q_j$ contains $j$ more powers of $\alpha$ than of $x$ so that 
Eq.~(\ref{bafour}) embodies a natural power-counting of threshold 
logarithms.  

The functional form of $E_{ij}$ for $t\bar{t}$ production is identical to that 
for $l\bar{l}$ production, except for the identification of the two separate 
channels, denoted by the subscript $ij$.  However, only the {\it leading} 
threshold corrections are universal. Final-state gluon radiation as well as
initial-state/final-state interference effects produce sub-leading logarithmic 
contributions that differ for processes with different final states.  Among
all $\{Q_j\}$ in Eq.~(\ref{bafour}), only the very leading one is universal.
This is the linear term in $P_1$ contained in $Q_0$, that turns out to be 
$P_1$ itself.  Since we intend to resum only the universal leading logarithms, 
we retain only $P_1$.  Hence, Eq.~(\ref{bafour}) can be integrated explicitly, 
and the resummed version of Eq.~(\ref{bone}) is
\begin{equation}
\hat{\sigma}_{ij}^{R;pert}(\eta,m^2)=\int_{z_{min}}^{z_{max}}dz
{\rm e}^{E_{ij}(\ln({1\over 1-z}),\alpha)}
\hat{\sigma}_{ij}'(\eta,m^2,z).\label{bthreep}
\end{equation}

The upper limit of integration in Eq.~(\ref{bthreep}) is set by the boundary
between the perturbative and non-perturbative regimes.  An intuitive
definition of the perturbative region, where inverse power terms are
unimportant, is provided by the inequality 
$\Lambda_{QCD} \over{(1-z)m}$ $\le 1$.
This inequality is identical to the expression in moment space, 
Eq.~(\ref{tseven}),  with the identification $n = \ln {1 \over 1-z}$.  In
momentum space, the same condition is 
realized by the constraint that all $\{Q_j\},\ j\ge 1$ be small 
compared to $Q_0$.  From the explicit expressions\cite{ref:edpapert} for 
the $\{Q_j\}$, one may show that this constraint corresponds to 
\begin{equation}
P_1\left(\ln\left({1\over 1-z}\right),\alpha\right)<1\ .
\label{pertmom}
\end{equation} 
Equation~(\ref{pertmom}) is equivalent to the requirement that terms that
are subleading according to perturbative power-counting are indeed subleading
numerically; Eq.~(\ref{pertmom}) is the essence of perturbation theory in this
context.  

As remarked above, we accept only the perturbative content of
principal-value resummation, and our cross section is evaluated accordingly.  
Specifically, we use Eq.~(\ref{bthreep}) with the upper limit of integration, 
$z_{max}$, calculated from Eq.~(\ref{pertmom}).  The upshot is an effective 
threshold cutoff on the integral over the scaled subenergy variable
$\eta$, but one that is calculable, {\it not} arbitrary.  
Our perturbative resummation probes the threshold down to the point
$\eta\ge \eta_0 =(1-z_{max})/2 $.  Below this value, perturbation theory, 
resummed or otherwise, is not to be trusted.  For the
top mass $m$ = 175 GeV, we determine that the perturbative regime is
restricted to $\eta \geq$ 0.007 for the $q{\bar q}$ channel and 
$\eta \geq$ 0.05 for the $gg$ channel.  These numbers may be converted to more 
readily understood values of the subenergy above which we 
judge our perturbative approach is valid:  at $m$ = 175 GeV, these are 1.22 GeV 
above the threshold in the $q{\bar q}$ channel and 8.64 GeV 
above the threshold in the $gg$ channel. The difference reflects the larger
color factor in the $gg$ case.  A larger color factor makes the 
non-perturbative region larger.  (One could attempt to apply 
Eq.~(\ref{bthreep}) all the way to $z_{max} = 1$, i.e., to $\eta =$ 0, but 
one would then be using a {\it model} for non-perturbative
effects, the one suggested by PVR, below the region justified
by perturbation theory.) 

It is useful to comment on the differences between our approach to resummation 
and another method published recently by Catani $\it {et~al}$ 
(CMNT)\cite{ref:catani}.  We both use the same universal leading-logarithm 
expression in moment space, but differences occur after the transformation to
momentum space.  In this paper, we set aside comments on mathematical aspects 
of their procedure and focus instead on phenomenological issues of interest.  
As remarked above, the Mellin transformation generates subleading terms in 
momentum space.  CMNT choose to retain all of these inasmuch as they perform 
the Mellin inversion numerically.  Instead, in keeping with the fact that
subleading logarithmic terms are not universal, we retain only the universal
leading logarithm terms in momentum space, and we restrict our phase
space integration to the region in which the subleading terms would not be 
numerically significant regardless.  The differences in the two approaches can 
be stated more explicitly if we examine the perturbative expansion of the
kernel ${\cal H}^{R}_{ij}(z,\alpha)$, Eq.~(\ref{bafour}).  If, instead of 
restricting the resummation to the universal leading logarithms only, we were 
to use the full content of Eq.~(\ref{bafour}), we would arrive at an 
analytic expression that is equivalent to the numerical inversion of CMNT, 
\begin{equation}
{\cal H}^{R}_{ij} \simeq 2\alpha C_{ij} 
\biggl[\ell n^2 (1-z) - 2\gamma_E \ell n (1-z)\biggr] + {\cal O}(\alpha^2).
\label{padovao}
\end{equation} 
In terms of this expansion, in our work we retain only the leading term 
$\ell n^2 (1-z)$ at order 
$\alpha$, but CMNT retain both this term and the subleading term 
$ - 2\gamma_E \ell n (1-z)$.  We judge that it is not justified to keep the
subleading term for three reasons: it is not universal; it is not the same 
as the subleading term in the exact ${\cal O}(\alpha^3)$ calculation; and it
can be changed arbitrarily if one elects to keep non-leading terms in moment
space.  As a practical matter, the subleading term is negative and numerically
significant.  In the $q\bar{q}$ channel at $m=175$ GeV and ${\sqrt S}=1.8$ TeV, 
its inclusion eliminates more than half of the contribution from the leading 
term.  In our view, the presence of numerically significant subleading 
contributions means that non-universal structures are not under control and 
begs the question of consistency\cite{ref:crit}.  A further justification for 
the retention of only the leading term is that it approximates the exact 
next-to-leading order result well, as shown in Fig.~1(a).  The choice made by 
CMNT reproduces only one-third of the exact next-to-leading order result.  The 
influence of subleading terms is amplified at higher 
orders where additional subleading structures occur in the CMNT 
approach with significant numerical coefficients proportional to $\pi^2$, 
$\zeta(3)$, and so forth.  Further comments about the different results 
in the two approaches are reserved to our discussion of predictions for 
cross sections at ${\sqrt S}=1.8$~TeV.

\section{Physical cross section}

In order to achieve the best accuracy available we wish to include in 
our predictions as much as is known theoretically.  
Our final resummed partonic cross section can therefore be 
written\cite{ref:edpapero,ref:edpapert}
\begin{equation}
\hat \sigma^{pert}_{ij}(\eta, m^2,\mu^2)=
\hat \sigma^{R;pert}_{ij}(\eta,m^2,\mu^2)-
\hat \sigma^{(0+1)}_{ij}(\eta,m^2,\mu^2)\Bigg|_{R;pert}+
\hat \sigma^{(0+1)}_{ij}(\eta,m^2,\mu^2)\ . 
\label{fthree}
\end{equation}
The second term is the part of the partonic cross section up to one-loop that is
included in the resummation, while the last term is the exact one-loop 
cross section\cite{ref:dawson}.
\begin{figure}
\vspace*{13pt}
%\leftline{\hfill\vbox{\hrule width 5cm height0.001pt}\hfill}
{\hskip 2.0cm}\hbox{\epsfxsize8.6cm\epsffile{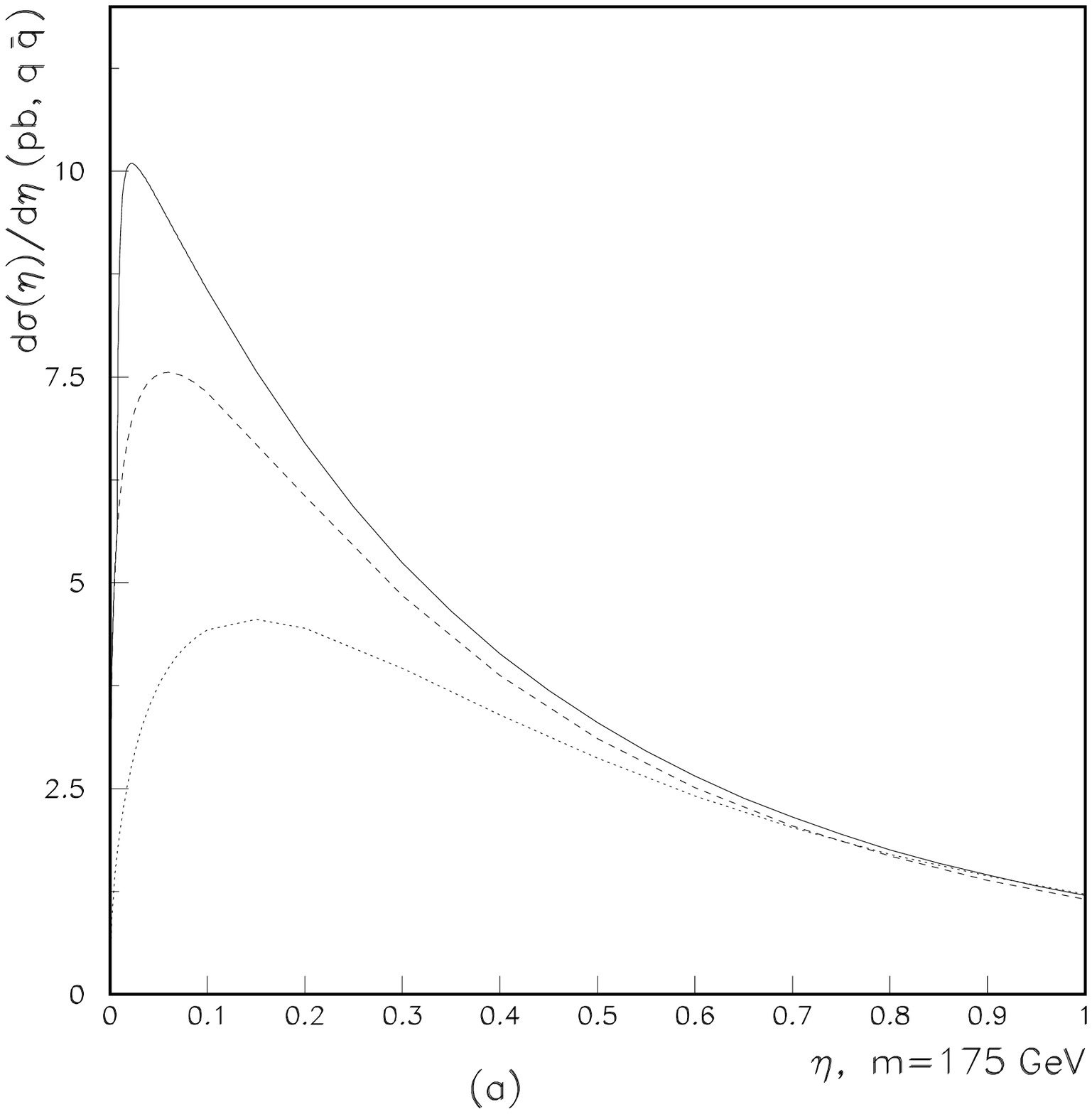}{\hskip -1.0cm}
\epsfxsize8.6cm\epsffile{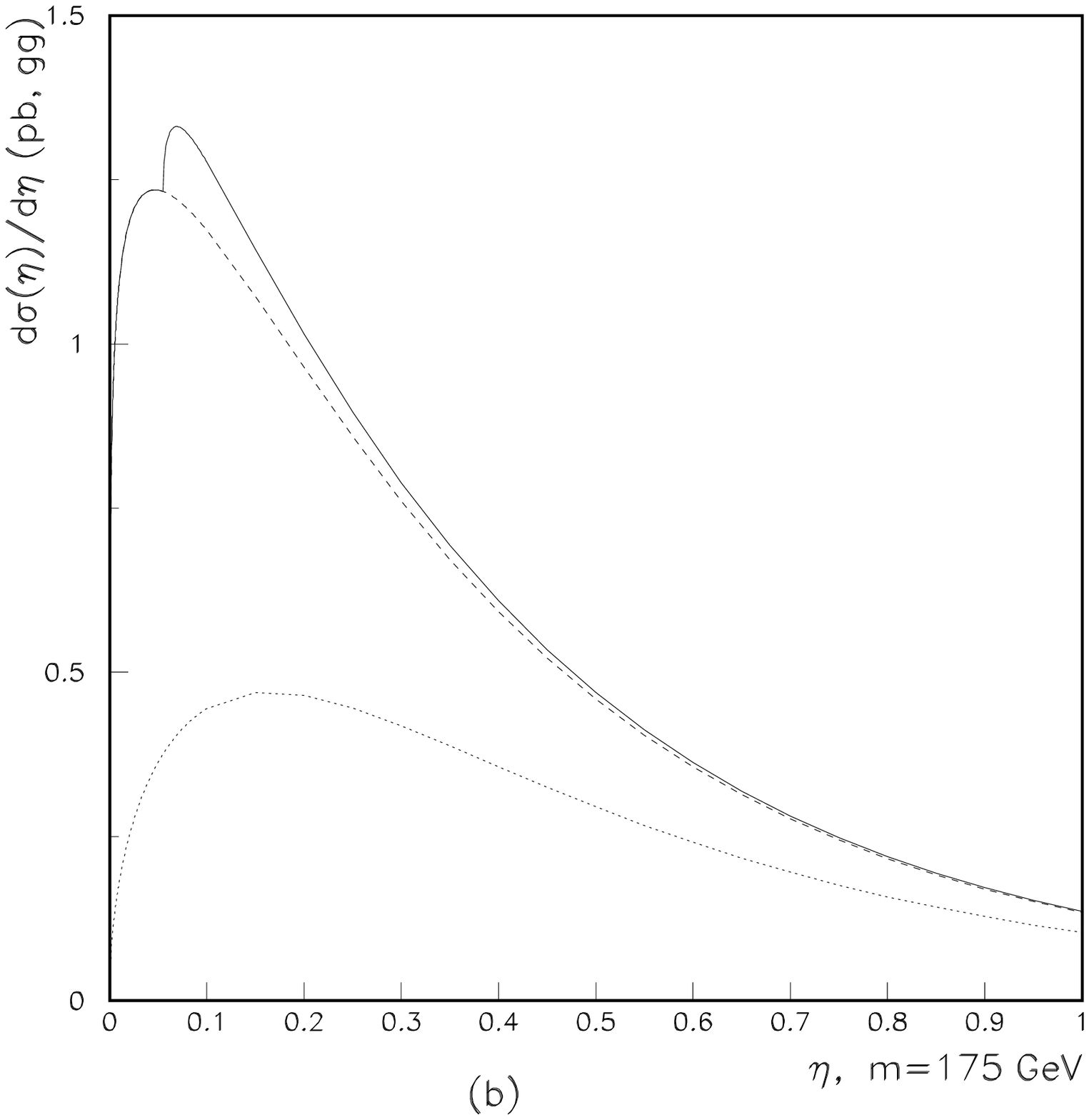}}
%\vspace*{1.4truein}             %ORIGINAL SIZE=1.6TRUEIN x 100% - 0.2TRUEIN
%\leftline{\hfill\vbox{\hrule width 5cm height0.001pt}\hfill}
\fcaption{Differential cross section $d\sigma/d\eta$ in the
${\overline{\rm MS}}$-scheme for the (a) $q\bar{q}$  and (b) $gg$  channels:
Born (dotted),  NLO (dashed) and resummed (solid).}
%\label{fig:fone}
\end{figure}
To obtain physical cross sections, we insert Eq.~(\ref{fthree}) into 
Eq.~(\ref{feleven}), and we integrate over $\eta$.  Other than the top mass,
the only undetermined scales are the QCD factorization and renormalization 
scales.  We adopt a common value $\mu$ for both, and we vary this scale over 
the interval $\mu/m\in\{0.5,2\}$ in order to evaluate the theoretical 
uncertainty of the numerical predictions.  We use the CTEQ3M parton 
densities\cite{ref:cteq}.

A quantity of phenomenological interest is the differential cross section 
${d\sigma_{ij}(S,m^2,\eta)\over d\eta}$.  Its integral over $\eta$ is the 
total cross section.  In Fig.~2 we plot these distributions 
for $m=175$ GeV, ${\sqrt S}=1.8$ TeV, and $\mu=m$.   The full range of $\eta$
extends to 25, but we display the behavior only in the near-threshold region
where resummation is important.  We observe that, at the energy of the 
Tevatron, resummation is significant for the $q\bar{q}$ channel and less so 
for the $gg$ channel.  In Fig.~1(a), the dotted curve shows that our final 
resummed cross section in the $q\bar{q}$ channel, after integration over all
$\eta$, lies about half-way between
the cross sections obtained from the near-threshold leading logarithms at
orders ${\cal O}(\alpha^3)$ and ${\cal O}(\alpha^4)$.  
\begin{figure}
\vspace*{13pt}
%\leftline{\hfill\vbox{\hrule width 5cm height0.001pt}\hfill}
{\hskip 2.0cm}\hbox{\epsfxsize8.6cm\epsffile{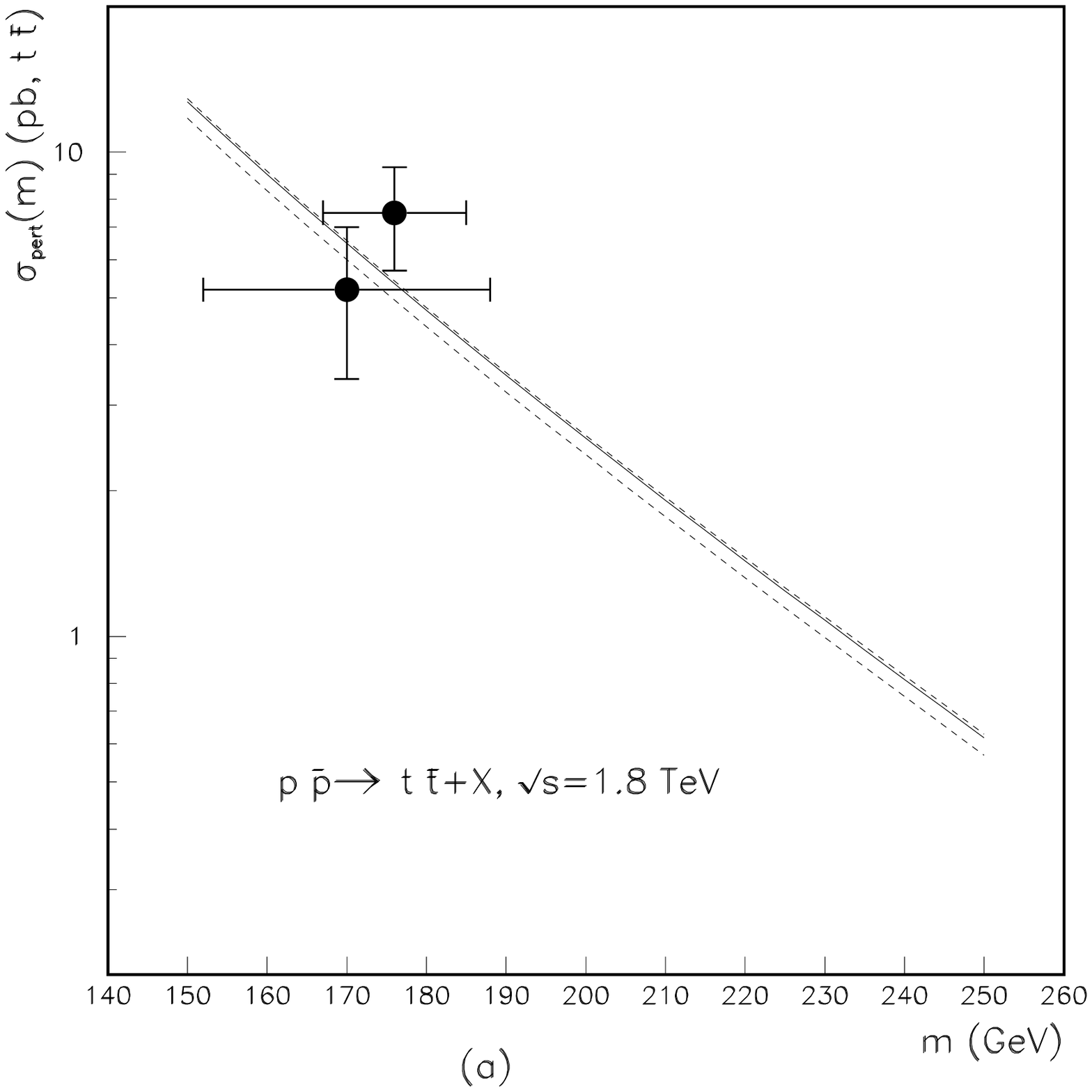}{\hskip -1.0cm}
\epsfxsize8.6cm\epsffile{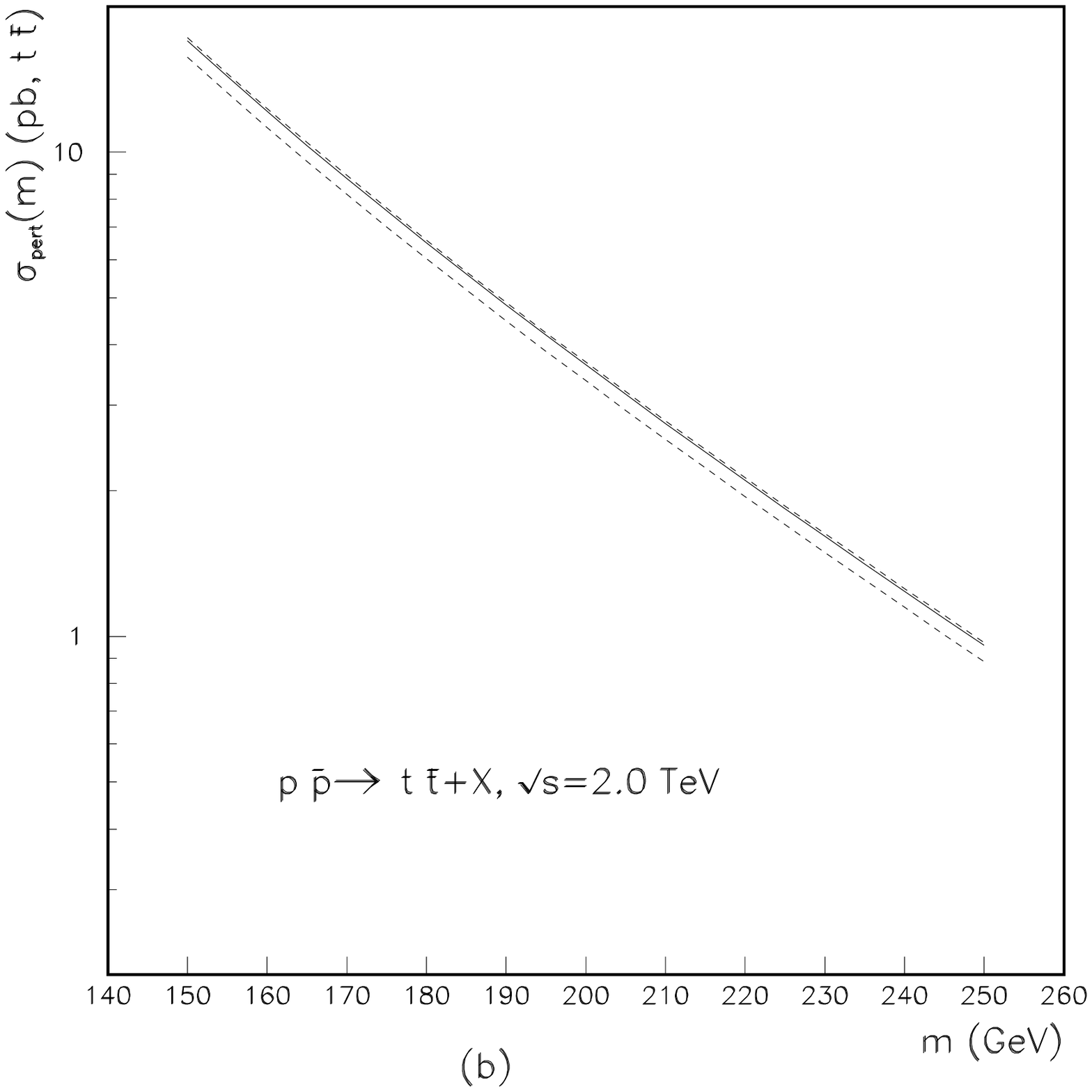}}
%\vspace*{1.4truein}             %ORIGINAL SIZE=1.6TRUEIN x 100% - 0.2TRUEIN
%\leftline{\hfill\vbox{\hrule width 5cm height0.001pt}\hfill}
\fcaption{Inclusive cross section for top quark production in the
${\overline{\rm MS}}$-scheme. The dashed curves show our perturbative
uncertainty band, while the solid curve is our central prediction:
(a) $\sqrt{S}=1.8$ TeV  and (b)  $\sqrt{S}=2$ TeV.}
%\label{fig:fone}
\end{figure}

We show our total $t\bar{t}$-production cross section as a function of top mass 
in Fig.~3.  The central value of our predictions is obtained with the 
choice $\mu/m=1$, and the lower and upper limits are  the maximum 
and minimum of the cross section in the range of the hard scale 
$\mu/m\in\{0.5,2\}$.  At $m =$ 175 GeV, the full width of the uncertainty band 
is about 10\%\ .  We consider that the variation of the cross section over 
the range $\mu/m\in\{0.5,2\}$ provides a good overall estimate
of uncertainty.  For comparison, we note that over the 
same range of $\mu$, the strong coupling strength $\alpha$ varies by 
$\pm10$\%\ at $m$ = 175 GeV.  Our prediction of Fig.~3(a) is in agreement with 
the  data\cite{ref:cdfdz}.  We find
$\sigma^{t\bar{t}}(m=175\ {\rm GeV},\sqrt{S}=1.8\ {\rm TeV})=
5.52^{+0.07}_{-0.42}\ pb$.  Our cross section is larger than the 
next-to-leading order value by about $9\%$.  Using a different choice of parton 
densities\cite{ref:mrsa}, we find a 4\%\ difference in the central value of our 
prediction\cite{ref:edpapero} at $m =$ 175 GeV.  A comparison of the 
predictions\cite{ref:edpapert} in the 
$\overline{\mbox{MS}}$ and DIS factorization schemes also shows a modest
difference at the level of $4\%$.

In Fig.~3(b) we present our predictions for an upgraded Tevatron operating
at $\sqrt{S}=2$ TeV.  We predict 
$\sigma^{t\bar{t}}(m=175\ {\rm GeV},\sqrt{S}=2\ {\rm TeV})=
7.56^{+0.10}_{-0.55}\ pb$.  The 2 pb increase in the predicted top quark cross 
section over its value at $\sqrt{S}= $1.8 TeV is about a 37\%\ gain.

Two other groups have published predictions for the cross section.  At
$m=175\ {\rm GeV}$ and $\sqrt{s}=1.8\ {\rm TeV}$, the three values are:
$\sigma^{t\bar t}$({\rm BC}\cite{ref:edpapero,ref:edpapert}) = 
$5.52^{+0.07}_{-0.42}$ pb;
\linebreak
$\sigma^{t\bar t}$({\rm LSvN}\cite{ref:laeneno}) = $4.95^{+0.70}_{-0.40}$ pb; 
and  
$\sigma^{t\bar t}$({\rm CMNT}\cite{ref:catani}) = $4.75^{+0.63}_{-0.68}$ pb.  
From the purely numerical point of view, all three predictions agree 
within their estimates of theoretical uncertainty.  However, the resummation 
methods differ, as discussed above, the methods for estimating the 
uncertainties differ, and different parton sets are used.  Comparing with 
LSvN\cite{ref:laeneno},  we find that our central values are 
$10-14\%$ larger, and our estimated theoretical uncertainty is 
$9-10\%$ compared with their $28\%-20\%$.  Our Born cross section, however, 
is about $3-5\%$ larger than the LSvN Born cross section due to the different 
parton distributions used in the two calculations.  Both the central value and
the band of uncertainty of the LSvN predictions are sensitive to their
undetermined infrared cutoffs.  To estimate theoretical uncertainty, 
we use the standard $\mu$-variation, whereas LSvN obtain their uncertainty
primarily from variations of their cutoffs.  CMNT calculate a central value of 
the resummed cross section (also with $\mu/m = 1$) that is less than 
$1\%$ above the exact next-to-leading order
value.  As explained earlier, the suppression of the effects of 
resummation arises from the retention by CMNT of numerically significant
non-universal subleading logarithmic terms.  Indeed, if the subleading 
term $ - 2\gamma_E \ell n (1-z)$ is discarded in Eq.~(\ref{padovao}), the
residuals $\delta_{ij}/\sigma_{ij}^{NLO}$ defined by CMNT\cite{ref:catani} 
increase from 
$0.18\%$ to $1.3\%$ in the $q\bar{q}$ production channel and from $5.4\%$ to 
$20.2\%$ in the $gg$ channel\cite{ref:MLMPN}.  After addition of the two 
channels, the total residual $\delta/\sigma^{NLO}$ grows from the negligible 
value 
of about $0.8\%$ cited by CMNT to the value $3.5\%$.  While still smaller than 
the increase of about $9\%$ that we obtain, the increase of $3.5\%$ vs. $0.8\%$ 
shows the substantial influence of the subleading logarithmic terms retained
by CMNT.  We explain above the reasons that we judge that our method 
is preferable theoretically. 

Turning to $pp$ scattering at the energies of the Large Hadron Collider (LHC)
at CERN, we note a few significant differences from $p\bar{p}$ scattering at 
the energy of the Tevatron.  The dominance of the $q {\bar q}$ production
channel is replaced by $g g$ dominance at the LHC.  Owing to the much larger 
value of $\sqrt{S}$, the near-threshold region in the subenergy variable is 
relatively less important, reducing the significance of initial-state soft
gluon radiation.  Lastly, physics in the region of large subenergy, where 
straightforward next-to-leading order QCD is also inadequate, becomes
significant for $t\bar{t}$ production at LHC energies.  Using the approach
described in this paper, we estimate
$\sigma^{t\bar{t}}(m=175\ {\rm GeV},\sqrt{S}=14\ {\rm TeV})= $ 760 pb.
\newpage
\section{Discussion and Conclusions}

In this paper, we summarize the calculation of the inclusive cross section
for top quark production in perturbative QCD, including the resummation of
initial-state gluon radiation to all orders.  The advantages of the
perturbative resummation method\cite{ref:edpapero,ref:edpapert} are that 
there are no arbitrary infrared cutoffs, and there is a well-defined 
perturbative region of applicability where subleading logarithmic terms are
numerically suppressed.  Our theoretical analysis shows that perturbative 
resummation without a model for non-perturbative behavior is both  possible and 
advantageous.  In perturbative resummation, the perturbative region of phase
space is separated cleanly from the region of non-perturbative behavior.
The former is the region where large threshold corrections exponentiate
but behave in a way that is {\it perturbatively stable}.
The asymptotic character of the QCD perturbative series, including 
large multiplicative color factors, is flat, and excursions 
around the optimum number of perturbative terms does not create
numerical instabilities or intolerable scale-dependence.  Infrared
renormalons are far away from the stability plateau and, even though
their presence is essential for defining this plateau, they are 
of no numerical consequence in the perturbative regime.  Large color factors, 
which are multiplicative in the exponent, enhance the infrared renormalon
effects and contribute significantly to limiting the perturbative regime.

Our resummed cross sections are about $9\%$ above the next-to-leading order
cross sections computed with the same parton distributions. The scale 
dependence of our cross section is fairly flat, resulting in a $9-10\%$ 
theoretical uncertainty.  This variation is smaller than the corresponding 
dependence of the next-to-leading cross section, as should be expected.
In recent papers\cite{ref:catani}, 
the authors state that the increase in cross section they find with 
their resummation method is of the order of $1\%$ over next-to-leading order.  
The numerical difference in the two approaches 
boils down to the treatment of the subleading logarithms, which can 
easily shift the results by a few percent, if proper care is not
taken. Our approach includes the universal leading
logarithms only while theirs includes non-universal subleading
structures which produce the suppression they find. In our opinion,  
their treatment of the subleading structures is not correct. 
The issue is one to be decided by the theory
community; it is not one for experimental resolution.

Our theoretical analysis and the stability of our cross sections under $\mu$
variation provide confidence that our perturbative resummation procedure 
yields an accurate calculation of the inclusive top quark cross section at 
Tevatron energies and exhausts present understanding of the perturbative 
content of the theory. Our prediction agrees with data, within the large 
experimental uncertainties. 

Extending our calculation to much larger values of $m$ than shown in Fig.~3,
we find that resummation in the principal $q\bar{q}$ channel produces 
enhancements over the next-to-leading order cross section of $21\%$, $26\%$, 
and $34\%$, respectively, at $m =$ 500, 600, and 700 GeV.  The reason for the
increase of the enhancements with mass at fixed energy is that the threshold 
region becomes increasingly dominant.  Since the $q\bar{q}$ 
channel also dominates in the production of hadronic jets at very large values 
of transverse momenta, we suggest that on the order of $25\%$ of the excess
cross section reported by the CDF collaboration\cite{ref:cdfjets} may well be 
accounted for by resummation.

This work was supported by the U.S. Department of Energy, Division of High 
Energy Physics, Contract No. W-31-109-ENG-38.


\begin{thebibliography}{9}
\bibitem{ref:crit}
It is difficult professionally to critique in Italy the work of four
esteemed Italian theorists, and it is uncomfortable personally as well.  
Stefano, Michelangelo, Paolo, and Luca are friends of many years.  They 
believe they are correct, and we are confident that we are.  Time will tell.
\end{thebibliography}
\end{document}